\documentclass[article]{IEEEtran}
\IEEEoverridecommandlockouts
\usepackage{cite}
\usepackage{amsmath,amssymb,amsfonts}
\usepackage{algorithmic}
\usepackage{graphicx}
\usepackage{textcomp}
\usepackage{xcolor}
\usepackage{xspace}
\usepackage{flushend}
\def\BibTeX{{\rm B\kern-.05em{\sc i\kern-.025em b}\kern-.08em
    T\kern-.1667em\lower.7ex\hbox{E}\kern-.125emX}}
\begin{document}

\IEEEoverridecommandlockouts
\IEEEpubid{\makebox[\columnwidth]{\hfill}}

\newcommand\WK[1]{\texttt{\textbf{#1}}}
\newcommand\wasm[1]{\texttt{\textbf{#1}}\xspace}
\newcommand\code[1]{\texttt{\textbf{#1}}\xspace}
\newcommand\arch[1]{\texttt{\textbf{#1}}\xspace}
\newcommand\engine[1]{\texttt{\textbf{#1}}\xspace}
\newcommand\ourengine{\texttt{\textbf{wizard}}\xspace}
\newcommand\OurEngine{\texttt{\textbf{Wizard}}\xspace}
\newcommand\OurJit{\texttt{\textbf{Wizard-SPC}}\xspace}
\newcommand\OurInt{\texttt{\textbf{Wizard-INT}}\xspace}
\newcommand\OurEngineLong{Wizard Research Engine\xspace}
\newcommand\X{$\times$\xspace}
\newcommand\NumCompilers{six\xspace}
\newcommand\NumCompilersMinusOne{five\xspace}
\newcommand\B[1]{\textbf{#1}}
\newcommand\C[1]{\textbf{#1}}

\title{Whose baseline compiler is it anyway?
}

\author{\IEEEauthorblockN{Ben L. Titzer} \\
\IEEEauthorblockA{\textit{Software and Societal Systems Department} \\
\textit{Carnegie Mellon University}\\
Pittsburgh, USA \\
btitzer@andrew.cmu.edu}
}


\maketitle

\thispagestyle{plain}
\pagestyle{plain}

\begin{abstract}
  Compilers face an intrinsic tradeoff between compilation speed and code quality.
  The tradeoff is particularly stark in a dynamic setting where JIT compilation time contributes to application runtime.
  Many systems now employ multiple compilation \textit{tiers}, where one tier offers fast compile speed while another has much slower compile speed but produces higher quality code.
  With proper heuristics on when to use each, the overall performance is better than using either compiler in isolation.
  At the introduction of WebAssembly into the Web platform in 2017, most engines employed optimizing compilers and pre-compiled entire modules before execution.
  Yet since that time, all Web engines have introduced new ``baseline'' compiler tiers for Wasm to improve startup time.
  Further, many new non-web engines have appeared, some of which also employ simple compilers.
  In this paper, we demystify single-pass compilers for Wasm, explaining their internal algorithms and tradeoffs, as well as providing a detailed empirical study of those employed in production.
  We show the design of a new single-pass compiler for a research Wasm engine that integrates with an in-place interpreter and host garbage collector using value tags, while also supporting flexible instrumentation.
  In experiments, we measure the effectiveness of optimizations targeting value tags and find, somewhat surprisingly, that the runtime overhead can be reduced to near zero.
  We also assess the relative compile speed and execution time of \NumCompilers baseline compilers and place these baseline compilers in a two-dimensional tradeoff space with other execution tiers for Wasm.
\end{abstract}


\IEEEpeerreviewmaketitle

\section{Introduction}
Software virtual machines (VMs) provide a way to execute a \emph{guest} programming language, instruction-set architecture, or bytecode format on a different \emph{host} machine.
VMs employ a variety of execution strategies that balance memory consumption, startup time, and peak performance.
In settings where loading or generating code at runtime is possible, new code can ``appear from nowhere'', and purely ahead-of-time translation is not possible.
This leaves such virtual machines with the option to employ an interpreter or a dynamic compiler.

\subsection{WebAssembly}

First appearing in major Web Browsers in 2017, WebAssembly~\cite{WasmPldi} (or \emph{Wasm}) is a bytecode format designed to offer portable native-level performance and software fault isolation via efficient in-process sandboxing.
Wasm is a machine-independent but machine-level compilation target that can be executed on modern CPUs with very low overhead.
It has allowed an explosion of new, powerful Web applications and capabilities, such as desktop applications like AutoCAD~\cite{WasmAutoCAD} and Photoshop~\cite{PhotoshopWeb}, video conference acceleration~\cite{WasmBlur}, real-time audio processing for echo reduction~\cite{AmazonEcho}, and many others.
Many of these are made possible by recompiling (potentially millions of lines of) legacy C/C++ code using standard toolchains that now support Wasm as a target.

WebAssembly is the first example of a major language that has employed formal specification and verification from design inception.
With a fully-formalized specification~\cite{WasmSpec} and machine-checked proof of type safety~\cite{WasmMechSpec}, it offers the most rigorously-specified compilation target to date, making it the most robust option for strongly isolating untrusted code on the Web or in-kernel.
In the literature, Wasm has inspired a number of exciting directions in Web research~\cite{DuckDB}~\cite{CTWasm}, verification research~\cite{MSWasm}~\cite{WasmSandboxing}, systems research~\cite{HFI}, cloud and edge computing~\cite{ServerlessFunctions}~\cite{SledgeWasm}~\cite{WasmIndMach}, and PL research~\cite{WasmK}.
~\nocite{EBPFCVE}

\subsection{Execution Strategies for Dynamic Code}
\B{Interpreters.}
An interpreter executes a program by examining \emph{data} that represents guest code.
Strict interpreters can execute any given input program without generating new machine code\footnote{Some interpreters may generate machine code stubs or, e.g. per-signature helper routines, but don't translate guest code directly to machine code.}.
Interpreters have the advantage of little or no up-front processing of the program is required and can often execute the code directly from the disk or wire format, saving both startup time and memory.
Interpreters also excel at debugging and introspecting execution states, as they often directly implement the state abstractions of their respective code format, such as an operand stack.
However, dispatch overhead means interpreters can never match the performance of compiled code in the long run.

\B{Baseline JIT compilers.}
VMs have deployed \emph{dynamic translation} to machine code as far back as LISP in 1960.
Often called just-in-time (JIT) compilation, a dynamic compiler generates new machine code at runtime that behaves equivalently to the interpreter's semantics, but is much faster.
A \emph{baseline} compiler is designed to generate machine code as fast as possible, forgoing the use of an intermediate representation (IR).
The very first dynamic translators were baseline compilers, stamping out templates of the interpreter's logic for each guest instruction or AST node, one after another, thus neatly eliminating the interpreter dispatch loop.
Despite the simplicity of baseline compilers, execution time improvements of 3\X to 10\X are common.

\B{Optimizing JIT compilers.}
JITs in today's virtual machines are powerful, integrating many ideas from static compilers, employing state-of-the-art IRs and sophisticated optimization passes.
For example, TurboFan~\cite{TurboFan}, the optimizing compiler in V8, employs a program dependence graph (PDG) representation called the ``sea of nodes''~\cite{SeaOfNodes}, with two different but overlapping optimization pipelines, one for JavaScript, and one for WebAssembly.
Key optimizations employed by most modern optimizing JITs are inlining, load elimination, strength reduction, branch folding, loop peeling and unrolling, global code motion, instruction selection, and register allocation.

\subsection{Overview and Contributions}

This paper is about maximizing compile speed for WebAssembly.
It presents a new single-pass compiler design for a research engine and compares and contrasts it with other single-pass compilers and other tiers for Wasm execution.
This paper's contributions are:

\begin{itemize}
\item \B{A new baseline compiler}, \OurJit, designed for interoperability with in-place interpretation of Wasm in the \OurEngineLong, supporting full-fidelity instrumentation and debugging.
\item \B{Distillation} of the key designs for \NumCompilersMinusOne other Wasm baseline compilers that all share the same foundational abstract interpretation approach, yet are discussed nowhere in the literature.
\item \B{Novel value tag optimizations} that reduce their runtime cost nearly to zero, greatly simplifying runtime systems.
\item \B{Novel instrumentation optimizations} that support flexible instrumentation for dynamic analysis.
\item \B{Empirical evaluation} of baseline compiler optimizations in \OurEngine, including novel value tag optimizations.
\item \B{Multi-tier performance comparison} among interpreters, baseline compilers, and optimizing compilers for Wasm.
\end{itemize}

As description of fast Wasm baseline compilers do not yet appear in the literature, this paper first clarifies these designs by describing the basic abstract interpretation algorithm which they all share.
We then report on \OurJit, a new, state-of-the-art single-pass compiler for a research Wasm engine.
A novel design problem is integrating with an in-place interpreter to support full-fidelity debugging and instrumentation of Wasm code.
For evaluation, we compare \NumCompilers baseline compilers found in industry across a wide variety of benchmarks and place them in context with interpreters and optimizing compilers.

\section{Executing Wasm}

Wasm bytecode is organized into modules, with top-level functions containing instructions for a stack machine.
Wasm bytecode is unusual in that it has structured control-flow constructs like \wasm{block}, \wasm{if}, and \wasm{loop}.
Such constructs improve the compactness of the code format and the efficiency of the code validation algorithm.
A key property is that branches that target a \wasm{block} or \wasm{loop} must be nested inside the construct.
This leads to a natural notion of a ``control stack'' that allows the validator algorithm to immediately reuse any internal metadata for control constructs as soon as the construct is exited\footnote{It is believed, but has not been yet shown, that this representation is optimally efficient.}.
Another intentional design property is that all control-flow predecessors of a label (except \wasm{loop}) precede the label, enabling highly efficient single-pass forward data flow analysis via abstract interpretation.

Wasm now exhibits execution tiers of all three basic designs.
Interestingly, these appeared in the exact opposite order to most VMs.
Optimizing compilers for Wasm appeared first in Web engines, made possible by the engineering effort put into making JavaScript fast.
Later, Web Engines added baseline compiler tiers, as startup time became an issue for large Wasm modules.
Concurrently, non-Web compilers and interpreters started appearing.
Initially, interpreters employed rewriting of Wasm code to another representation, but recent work~\cite{FastWasmInterp} showed an in-place interpreter can be on par with rewriting interpreters.

\section{Single-pass Compilation of Wasm}

Single-pass compilers for Wasm are designed for compile speed and simplicity.
A single pass affords no time to build an intermediate representation of the code.
Instead, such compilers are limited to generating code for one (or a small number) of instructions at a time based on limited context accumulated from prior instructions.

\subsubsection{The Abstract-Interpretation Approach}

  \begin{figure*}
    \includegraphics[width=7.1in]{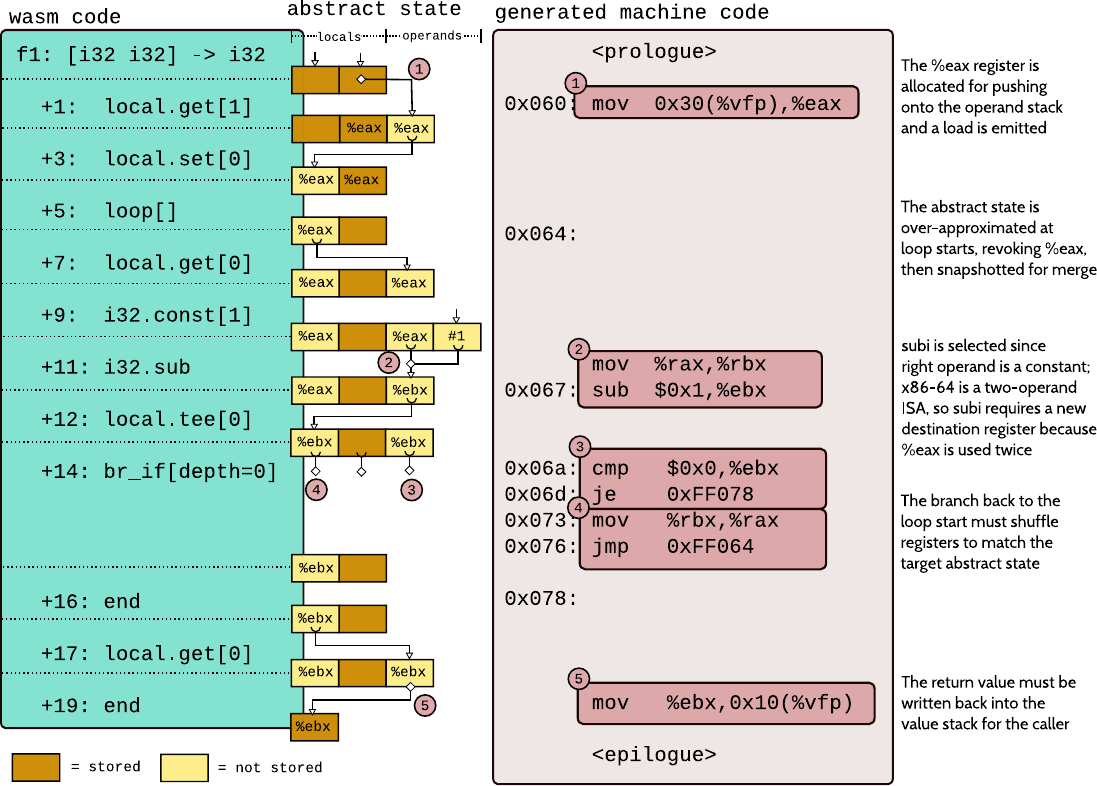}
    \caption{Illustration of single-pass compilation using abstract interpretation. (Actual code emitted by the Wizard single-pass compiler.)}
    \label{fig:compilation-diagram}
    \end{figure*}

In our study of Wasm compilers, we found that all single-pass compilers are simply variations on a basic abstract-interpretation approach that is similar to Wasm code validation\footnote{In fact, some baseline compilers in this study, like Liftoff, reuse parts of their validation algorithms to drive compilation.}.
Thus, by understanding this common approach, we can compare and contrast the variations and more easily understand the innovation represented by \OurJit.

Figure~\ref{fig:compilation-diagram} gives an example compilation of Wasm code using the common abstract-interpretation approach.
The abstract state consists of an \emph{abstract value stack} (shown), an \emph{abstract control stack} (not shown), and register allocation state (not shown).
Each local variable slot and operand stack slot in the abstract value stack has an \emph{abstract value} that can contain information such as:
\begin{itemize}
\item \B{stored} - tracks whether the slot has been stored into memory, and where,
\item \B{register} - the register, if any, which holds the value, and
\item \B{const} - the concrete value, if a constant.
\end{itemize}
Local variables representing parameters are initialized from the signature of the function and the calling convention.
In the example, argument values arrive from the caller in an explicit value stack stored in memory.
Declared local variables (not shown) are by Wasm semantics initialized to the default value for their respective type (i.e. a constant).
Not shown, the algorithm maintains an abstract control stack that tracks the nesting of control constructs such as \wasm{block} and \wasm{loop}.
Each construct has a \emph{label} which represents the place in the machine code where branches targeting it will jump.

After emitting a few machine instructions for the prologue, compilation procedes by examining each instruction in sequence.
Instructions that access locals (\wasm{local.get}, \wasm{local.set} and \wasm{local.tee}) manipulate the abstract state.
Depending on whether the local or top-of-stack is allocated to a register, the compiler may emit a load or store instruction, but often emits no code at all.
We'll see in the next section that variations in the abstract state of the compilers we examined impact the amount of moves generated.
For constant-generating instructions like \wasm{i32.const}, abstract values can model concrete values and avoid generating any code at all.
Of the \NumCompilers compilers we study, all but one model constants.

Control flow requires the compiler to manage \emph{snapshots} of the abstract state that represent the contents of registers and the stack at labels, i.e. merges in control flow.
All constructs except \wasm{loop} have their label at the end, which means that all branches to the label will be seen before the label itself.
For \wasm{loop}, absent any knowledge of the code in the loop body, compilers must over-approximate the abstract state before compiling the loop body, e.g. by assuming all slots could be modified on backedges.

A key design consideration in making a fast compiler is efficiently snapshotting the abstract state and merging states coming from multiple branches, since the abstract state can have tens of thousands of slots for large functions.
Different compilers we studied have different strategies, either making copy extremely cheap (i.e. \texttt{memcpy}), keeping a delta index, or tracking only a subset of slots and spilling the rest.
A nice benefit of Wasm's structured control flow is that the snapshot for a merge point can be deallocated as soon as a control construct is exited.
These considerations help avoid JIT bombs, which are small programs that exploit a non-linearity in the algorithmic complexity of a compiler as a form of denial-of-service attack~\cite{VerifierDos}.

The deceptively simple compilation approach is quite tricky to implement correctly and efficiently, but nevertheless yields surprisingly good code, as can be seen in the example.
Wasm's control flow design helps here; labels (other than \wasm{loop}) will have had all their predecessors visited before they are reached, allowing abstract interpretation to propagate constants to merge points in a single forward pass.
All-in-all, a single-pass compiler can peform:

\begin{itemize}
\item \B{register allocation} - if abstract values track register occupancy for each slot, codegen can elide code for most local accesses, often just updating the abstract state,
\item \B{constant-folding} - if abstract values track constants, codegen can compile-time evaluate side-effect free instructions, producing more constants,
\item \B{branch-folding} - if abstract values track constants, then branches whose input condition is a constant can be removed or compiled to unconditional jumps,
\item \B{strength-reduction} - if abstract values track constants, then some simple patterns such as \wasm{(i32.add x (i32.const 0))} can be reduced or eliminated,
\item \B{instruction selection} - if abstract values track register occupancy, codegen can select memory or register addressing modes, and if abstract values track constants, it can emit immediate-mode instructions such as \texttt{addi},
\item \B{avoid redundant spills} - if the abstract values track spill state, codegen can avoid repeated spills to the stack in subsequent instructions, and
\item \B{peephole optimization} - if codegen can peek one or more instructions ahead, it can combine multiple instructions, such as a compare and a branch.
\end{itemize}

From our study of baseline Wasm compilers, code generation for most opcodes (over 440 in Wasm today) is tedious and formulaic but not intrinsically difficult.
The crux of good single-pass compilation is two subtle things that require careful data structure design.
First, \B{managing the abstract state}, whose size is proportional to the locals and operand stack, must be done carefully and efficiently at all control flow points (branches, loops, and merges) to avoid (or at least mitigate) quadratic compilation time.
And second, abstract values should model constants \emph{and} registers, allowing \B{efficient forward-pass register allocation} so that most Wasm instructions use machine registers and avoid spills.
The two are intertwined; the abstract state of all compilers contains register assignments and must be checkpointed at control flow split points and merged at control-flow join points.

\section{Baseline Compiler Integration}

  \begin{figure*}
    \includegraphics[width=7.1in]{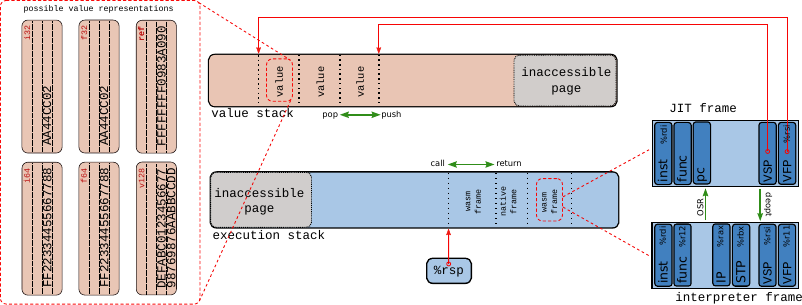}
    \caption{Execution frame and value stack layout for \OurInt and \OurJit.
      Both kinds of execution frames are the same number of machine words, allowing quick tier-up (OSR) and tier-down (deopt) by rewriting execution frames in place.
    The interpreter directly manipulates the value stack in memory, while JIT code only spills to the value stack when registers are exhausted and across calls.}
    \label{fig:jit-frame-diagram}
    \end{figure*}

  A JIT compiler in any virtual machine must integrate with other execution tiers and services such as debugging, instrumentation, and garbage collection.
  Thus, in a mature system, a JIT compiler becomes invisible, and users experience better performance with no loss of functionality.
  This becomes progressively more complicated with more execution tiers, as handoff between different types of code efficiently can involve very delicate machine code tricks.
  In this section we cover aspects of integrating \OurJit that motivated and are in turn constrained by \OurEngine's prior design decisions.
  
  \subsection{In-place Interpreter Integration}

  Prior to this work, the \OurEngineLong was an interpreter-based system with debuggability and introspection the main priorities.
  The in-place interpreter~\cite{FastWasmInterp} (hereafter referred to as \OurInt) executes Wasm code without rewriting, allowing tracing, profiling, debugging in terms of the original bytecodes and offsets.
  
  Relevant points on the \OurInt design are:
  \begin{itemize}
  \item Interpreter performance is competitive with production interpreters for Wasm.
  \item The value stack is explicitly emulated at runtime, including locals and operand stack values.
  \item Stack walking uses \emph{value tags} to precisely find GC roots (\wasm{externref} and Wasm GC objects).
  \item Users can insert \emph{probes} into bytecode locations which call back to instrumentation and implement tracing, debugging, and profiling in an extensible way.
  \end{itemize}

The last three points were addressed in \OurJit by 1) using an identical value stack and nearly-identical execution stack layout as \OurInt, 2) emitting (and optimizing) value tag stores in JITed code, and 3) emitting efficient callbacks to user code and intrinsifying key probe kinds.
  
\subsection{Value Stack and Execution Frame Layouts}

As we've seen, all single-pass compilers for Wasm use abstract interpretation to statically compute the operand stack height and approximate stack contents at every instruction in a function.
Some of the baseline compilers we studied \emph{reallocate} the storage of operand stack slots and locals to machine stack slots and registers, i.e. they scramble the stackframe layout.
Scrambling the stack creates a mapping problem for debugging and instrumentation: where are original values stored on the machine stack?
This metadata imposes a space cost, and is remarkably complex, tricky and error-prone.
In fact, of the \NumCompilersMinusOne previous compilers we studied, only two support introspection in their baseline compilers; the others just \emph{do not support debugging at all}.

\OurEngine's baseline compiler is meant to integrate with \OurInt that has an exact model of the value stack.
It does not scramble the stack, and moreover, uses a nearly identical execution frame layout between the interpreter and JITed code.
In Figure~\ref{fig:jit-frame-diagram}, we see the layout of execution frames in \OurEngine for the interpreter and JIT code.
Both use the same value stack representation for storing Wasm values and only differ in their native execution (machine stack) frames.
In particular, interpreter frames contain bytecode-level pointers (\wasm{IP}), a sidetable pointer (\wasm{STP}), and additional metadata.
When executing in the interpreter, more registers are needed to store these additional pointers, whereas in JIT code, only the value frame pointer (\wasm{VFP}), instance (\wasm{inst}), and memory base (not shown) are needed.
That leaves more registers to be allocated to compiled code.
While values are in registers, the value stack in memory may not be up-to-date.
At observable points like outcalls, JITed code simply writes values into the value stack in memory.
For stacktraces, instrumentation, and debugging, the current program counter (i.e. bytecode offset) can be recomputed from the machine code instruction pointer or explicitly saved into the execution stack.

The compatibility between the two frame layouts allows \OurEngine to \emph{tier-up} from \OurInt to baseline-compiled code (e.g. when a function is detected as \emph{hot}) by changing only the execution frame and jumping into JITed machine code.
Conversely, \OurEngine can \emph{tier-down} (for debugging or to support user instrumentation) by simply reconstructing \wasm{IP} and \wasm{STP} and jumping back into \OurInt.

\subsection{Value Tags versus Stackmaps for GC}

Wasm code can contain references to host objects as values of type \wasm{externref}.
In a host environment with precise garbage collection, the VM must find all roots, including those that may be in the Wasm value stack.
There are two basic strategies that allow the VM to distinguish references from non-references: \emph{stackmaps} or \emph{value tags}.
The primary difference between the two is that stackmaps are basically static and value tags are basically dynamic.

\textbf{Stackmaps.}
For JIT-compiled code, compilers often emit metadata called \emph{stackmaps} attached to the code which encodes how to find references in stack frames of JITed code.
Such metadata usually adds space proportional the size of JITed code, so it is often very compactly stored.
It is also notoriously hard to get right, as bugs in stack walking logic or errors in compressed metadata result in VM-level crashes that are insanely tedious to debug\footnote{Generally the least welcome type of GC+JIT bug.}.
Despite the added complexity and potential robustness problems of stackmaps, they have less dynamic cost, normally only used during GC.

\textbf{Value Tags.}
Value tags are an entirely dynamic strategy where values themselves contain the metadata that distinguishes references from non-references.
This metadata can be encoded in various ways, such as a tag bit, an indirection, a value range restriction, or often an additional byte or word, such as a tag byte or dynamic type information.
The possibilities for encodings varies with the kinds of values that are used to implement the guest language.
Value tags allow the VM to easily inspect a value anywhere in memory (such as GC scanning stacks for references) making it vastly simpler and more robust.
Another important advantage is that a JIT compiler may avoid stackmaps altogether, saving compile time and space.
A disadvantage is the dynamic cost, since tags require additional space and may introduce dynamic checks.

Of the Wasm engines in the wild, including the ones containing the \NumCompilers baseline compilers, none use value tags except \OurEngine.
These systems either do no precise garbage collection at all, removing the need for stackmaps, or they reuse the battle-tested stackmap logic of their host system, as is the case in all Web engines.
Since \OurEngine makes unusual choices here, we evaluate some of the tradeoffs specific to that design in the experimental section.

\begin{figure*}
  \begin{small}
  \begin{tabular}{||l | l | l | l | l||}
    \hline
    Name & Language & Year & Features & Description \\
    \hline
    \engine{wizeng-spc}  & Virgil & 2023 & \verb|MR K KF ISEL TAG MV| & The \OurEngineLong's single-pass compiler. \\
    \engine{wazero}      & Go     & 2022 & \verb| R                 | & An open-source engine written in Go~\cite{Wazero}. \\
    \engine{wasm-now}    & C++    & 2022 & \verb|MR K    ISEL       | & A research project using Copy\&Patch~\cite{CopyAndPatch} code generation. \\
    \engine{wasmer-base} & Rust   & 2020 & \verb| R K             MV| & The \engine{--singlepass} option of \engine{wasmer}~\cite{Wasmer}. \\
    \engine{v8-liftoff}  & C++    & 2018 & \verb|MR K    ISEL MAP MV| & The baseline Wasm compiler in \engine{V8}~\cite{Liftoff}. \\
    \engine{sm-base}     & C++    & 2018 & \verb|MR K    ISEL MAP MV| & The baseline Wasm compiler in \engine{Spidermonkey}~\cite{SpiderMonkey}. \\
 \hline
  \end{tabular}
  \end{small}
  \caption{WebAssembly baseline compilers used in this study.
    \texttt{MR} = multiple register allocation,
    \texttt{R} = register allocation,
    \texttt{K} = constant tracking,
    \texttt{KF} = constant-folding,
    \texttt{ISEL} = instruction selection,
    \texttt{TAG} = value tags,
    \texttt{MAP} = stackmaps,
    \texttt{MV} = multi-value, a Wasm feature where blocks can return (multiple) values.
  }
  \label{fig:baseline-tiers}
\end{figure*}

\textbf{Optimizing Value Tags.}
The dynamic cost of value tags can be reduced with compiler optimization.
While an optimizing compiler can use a sophisticated global register allocator to only store tags on spills, a baseline compiler cannot afford an IR.
Instead, we outline three optimizations for reducing the dynamic cost of value tags in a single-pass compiler.

\begin{itemize}
\item \textbf{\emph{lazy tagging}} of locals.
  Since Wasm is a typed bytecode, local variables have static types that do not change during the execution of a function.
  Thus the types of locals can determined from the first bytes of a function body.
  Instead of writing value tags at runtime, the stackwalker computes them on-the-fly by decoding local declarations in the original bytecode, needing no additional metadata.
\item \textbf{\emph{lazy tagging}} of operand stack.
  While the types of local variables of a Wasm function don't change during execution, the types of operand stack slots certainly can.
  With this optimization, the compiler omits tag stores for operand stack slots.
  Like lazy tagging for locals, types are reconstructed at stackwalking time, but this is more complicated than for locals, because the types could be different at each bytecode.
  That means storing additional metadata (basically a stackmap), or reconstructing them from the bytecode by revalidating the code.
\item \textbf{\emph{on-demand tagging}} using abstract interpretation.
  The default for \OurJit, value tag stores are only emitted by the compiler across possible observations (calls, traps, and instrumentation) and the abstract state tracks whether each slot has an up-to-date tag.
  Parameters are assumed to have their tags stored by the caller.
\end{itemize}

We evaluate these alternatives in the experiments section by comparing with the worst-case overhead (an implementation that always stores value tags at each instruction, exactly as an interpreter would do) and the best-case alternative of simply disabling value tags.
As we will see, on-demand tagging mostly eliminates tag overhead.

\subsection{Supporting and Optimizing Instrumentation}

Like \OurInt, \OurJit supports flexible instrumentation via local \emph{probes}, which are user callbacks that fire before a given instruction is executed.
\OurEngine exposes an API that allows user code called a \emph{monitor} to instrument Wasm modules as they are loaded.
Probes are written in the implementation language of engine and when fired can access both engine APIs and the state of the Wasm program, including the values in execution frames, memory, tables, etc.
In the interpreter, a probed instruction triggers a call to \OurEngine's runtime system which looks up attached probes and fires them, passing an opaque, lazily-allocated \emph{accessor object} that exposes methods to access to the frame's internal state.
Probes are supported transparently, and more efficiently, in \OurJit.
\OurJit inserts direct calls at probed instructions, saving indirection through the runtime by statically determining which probes to fire for which instruction.
\OurJit further optimizes certain types of probes by emitting specialized machine code, such as a direct increment of a counter's value or directly passing the top-of-value-stack to a probe, eliding the accessor object.

\section{Baseline Compiler Comparison}

We studied the implementation of \NumCompilers single-pass compilers for WebAssembly that employ the basic abstract-interpretation algorithm.
The table in Figure~\ref{fig:baseline-tiers} compares their designs in terms of features.
In particular, we find that both Web engine compilers (\engine{v8-liftoff} and \engine{sm-base}) implement GC with stackmaps, using the same metadata format as their optimizing compilers.
As discussed, \OurJit uses value tags, and the three remaining compilers \emph{do no GC}, because their host environment is not garbage-collected.
A key feature is \emph{multiple register allocation}, where the abstract state allows a register to be used for more than one slot.
This is more complex to track and merge efficiently, but experimental results show it significantly improves code quality.
All compilers except \engine{wazero} track constants.
Experiments also show that tracking constants measurably improves code quality, as it allows some local instruction selection.
Of the \NumCompilers, only \OurJit performs constant-folding and branch-folding, though our experiments show that it has marginal benefit for the benchmarks studied.

\section{Experiments}

This section details a number of experiments we conducted to evaluate \OurJit's optimizations and design choices, compare it against other baseline compilers, and place baseline compilers in context with other tiers.

\B{Benchmark Suites.} 
We use three different benchmark suites: PolyBenchC~\cite{PolyBenchC}, an often-used suite of numerical kernels, Libsodium~\cite{LibSodiumBench} a suite of cryptographic primitive benchmarks, and Ostrich~\cite{OstrichBench}.
Each of these suites consists of a number of \emph{line-items} comprised of different programs (28 for PolyBenchC, 39 for Libsodium, and 11 for Ostrich), each compiled into a separate Wasm module.

\subsection{Speedup over Interpreter}

\begin{figure}
  \includegraphics[width=3.5in]{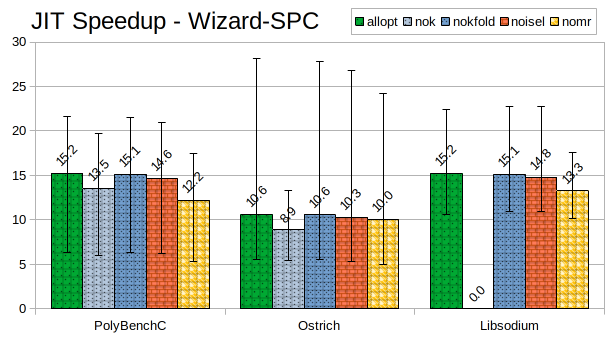}
  \caption{Execution time speedup of \OurJit over \OurInt (1\X $=$ same speed, 10 = 10\X faster, \emph{up} is better).}
  \label{fig:jit-speedup}
\end{figure}

Our first experiment evaluates speedup over \OurEngine's existing configuration with its in-place interpreter (\OurInt).
Here we focus on code quality by measuring the \emph{main execution time}, the time from the start of the program's main function until program exit.
This intentionally factors out VM startup and compilation time, pitting the interpreter speed against the speed of compiled code directly.
We study startup and compilation time in following experiments.

We evaluate five different optimization settings of \OurJit to assess the impact of each optimization.
\begin{itemize}
\item \B{allopt} - (default) all optimizations turned on.
\item \B{nok} - abstract values do not track constants, thus no constant-folding or instruction selection.
\item \B{nokfold} - no constant-folding or branch-folding.
\item \B{noisel} - no instruction selection, e.g. immediate modes.
\item \B{nomr} - no ``multi-register'' support; a register can cache at most one slot at a time.
  \end{itemize}

Figure~\ref{fig:jit-speedup} summarizes speedups across the three benchmark suites.
For each configuration, we run each benchmark line item 25 times, each time in a separate VM instance (9750 data points).
The height of each bar corresponds to the average speedup across line items in that suite.
Note the error bars are not measurement variance\footnote{
While there is significant variance amongst line items, measurements for a single line item are stable within a small variance.},
but variance amongst line items in that suite, i.e. the minimum and maximum average speedup for any line item.

From these results we can see that the compiled code runs between 5\X and 28\X faster than the interpreter for all line items, while suites averages are 10\X to 15\X.
From the \B{nok} configuration, we can see that disabling constant tracking in abstract interpretation has the most dramatic effect on code quality.
Disabling multiple register allocation (\B{nomr}) is significant, in some cases larger than disabling constant tracking.
Finally, disabling constant-folding (\B{nokfold}) and instruction selection (\B{noisel}) are small but measurable effects.

\subsection{Optimizations for Value Tags}

\begin{figure}
  \includegraphics[width=3.5in]{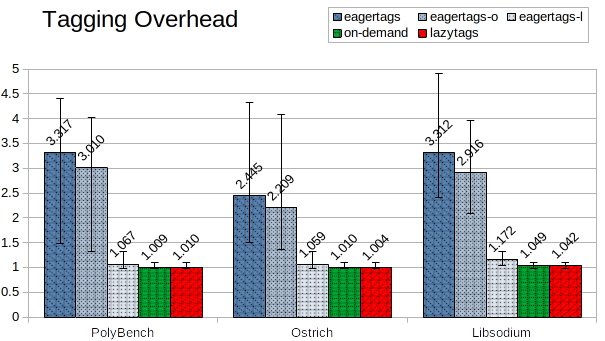}
  \caption{Execution time of \OurJit tagging configurations relative to a no-tagging configuration (1.0 = same speed as \engine{notags}, \emph{lower} is better).}
  \label{fig:tagging-overhead}
\end{figure}

Our second experiment in Figure~\ref{fig:tagging-overhead} compares design alternatives for \OurJit's support for value tags.
Using the same measurement methodology as the previous experiment, we measure relative main execution time of various tagging configurations.
Here, the baseline in the figure is no longer \OurInt but \B{notags}, where we disabled value tags altogether, including removing their space from the value stack.
The configurations tested here are:
\begin{itemize}
\item \B{eagertags} - ``eagerly'' store modified tags at every instruction.
\item \B{eagertags-o} - ``eagerly'' store tags for operand slots only.
\item \B{eagertags-l} - ``eagerly'' store tags for locals only.
\item \B{on-demand} - (default) store tags on-demand by tracking their state in abstract interpretation.
\item \B{lazytags} - store tags on-demand, but leave tagging of locals to the stack walker.
  \end{itemize}

Figure~\ref{fig:tagging-overhead} shows the average relative main execution time over the line-items in each benchmark suite.
As before, error bars represent the minimum and maximum of any line item in the respective suite.
We see that the eager-tagging imposes a 2.4\X - 3.3\X overhead on execution time.
By measuring eager-tagging of locals separately from the operand stack, we can attribute that overhead mostly to tagging of the operand stack\footnote{Which is to be expected, as the operand stack is where the action is!}.
We also see that the default \B{on-demand} tagging strategy almost completely eliminates the cost of value tags, within 0.9 - 4.9\% of the ideal \B{notags} configuration.
We can also see that \B{lazytags} can further reduce the tagging overhead of \B{on-demand}, statistically measurable, but the improvement is marginal, to 0.4 - 4.2\% on average.
Given that \B{lazytags} would imply design complexity to perform tagging in the stack walker, it was not productionized.

\subsection{Instrumentation optimizations}

\begin{figure}
  \includegraphics[width=3.5in]{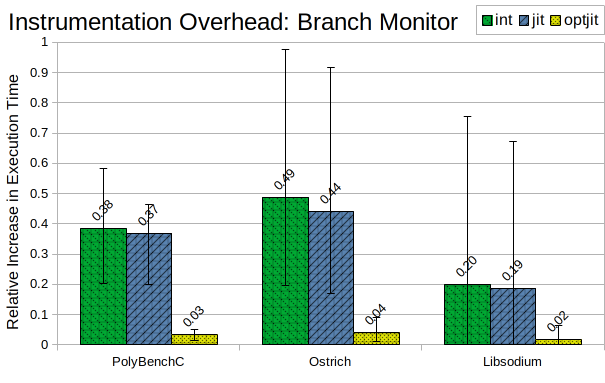}
  \caption{Probe overhead in \OurInt and \OurJit, reported as the increase in execution time relative to the interpreter. 
    (0.0 $=$ same speed, 1.0 = increase of 1\X the interpreter execution time; \emph{lower} is better). }
  \label{fig:probe-overhead}
\end{figure}

Our next experiment evaluates the effectiveness of \OurJit's optimizations targeting instrumentation.
In Figure~\ref{fig:probe-overhead}, we report measurements with the \emph{branch monitor}, a standard \ourengine tool that profiles the targets of all conditional branches using a local probe that reads the top-of-value-stack.
We show the increase in main execution time normalized to each line item's execution time on the interpreter, grouped by benchmark suite, and with error bars as before.
In interpreted mode (\textbf{int}), this monitor imposes a moderate 20-49\% average slowdown per suite.
Without optimization (\textbf{jit}), \OurJit simply emits calls to probe code which produces similar but slightly lower overhead.
It's similar because the overhead consists of runtime calls, the allocation of the accessor object, and accesses of the value stack through it, which are all in engine code.
In the (\textbf{optjit}) configuration however, \OurJit emits direct calls to the probe passing the top-of-stack value, skipping the runtime and accessor object allocation, which reduces overhead by approximately 10\X.
Of course JIT code runs 10-30\X faster than the interpreter, so renormalizing the data in Figure~\ref{fig:probe-overhead} to the JIT baseline, without optimizations the branch monitor slowdown is 5.4-9\X, which reduces to 42-77\% with optimization.

\subsection{Baseline shootout}

\begin{figure}
  \includegraphics[width=3.5in]{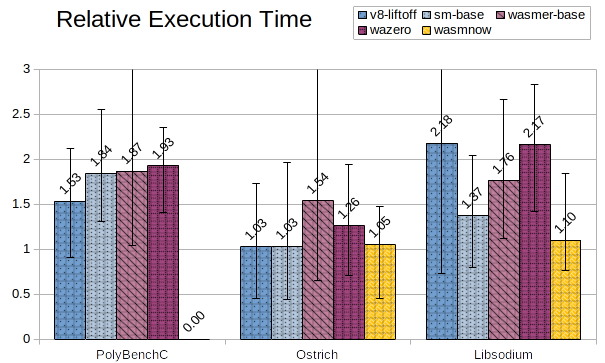}
  \caption{Relative execution time over \OurJit for other baseline compilers. (1.0 $=$ same speed, 2.0 = 2\X as long; \emph{lower} is better). }
  \label{fig:baseline-execution-time}
\end{figure}

\begin{figure}
  \includegraphics[width=3.5in]{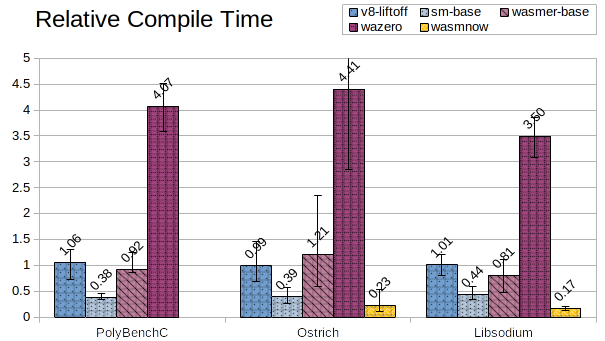}
  \caption{Relative compilation time over \OurJit for other engines in their baseline compiler configurations. (1.0 $=$ same speed, 2.0 = 2\X as long; \emph{lower} is better). }
  \label{fig:baseline-compile-time}
\end{figure}

\begin{figure}
  \includegraphics[width=3.5in]{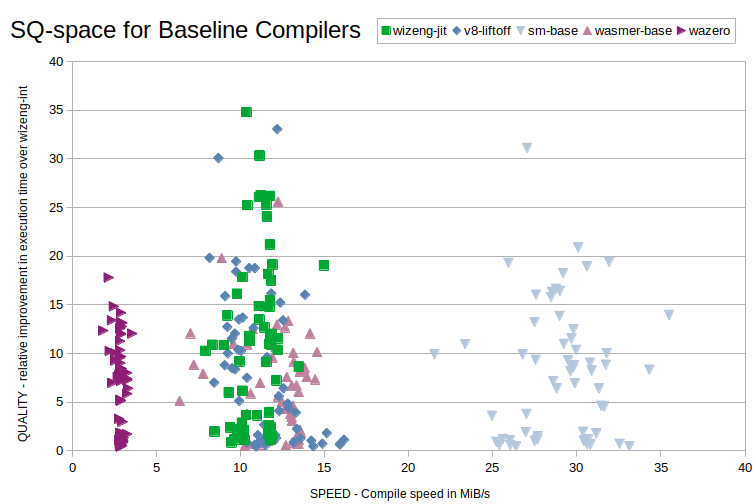}
  \caption{SQ-space comparison for baseline compilers. Quality is measured by the relative improvement in execution time over \OurInt. (1.0 $=$ same speed, 2.0 = 2\X as fast; \emph{up} and \emph{right} are better). }
  \label{fig:sq-space-baseline}
\end{figure}

Our next experiment compares the compile speed and code quality of baseline compilers listed in Figure~\ref{fig:baseline-tiers}.
To gather compile times, we instrumented each engine to measure and report the time taken to compile each module, as well the number of input Wasm code bytes.
We compute the compile time as the time taken \emph{per byte} of input code, which naturally normalizes across different function and benchmark sizes\footnote{and also controls for lazy compilation, though engines in this study were configured to eagerly compile modules when possible.}.
We normalize the results relative to \OurJit for each line item.

Figure~\ref{fig:baseline-compile-time} displays the results of measuring compile time.
The height of each bar represents the compile time per byte of input code normalized to \OurJit, averaged over the line items in each suite.
The error bars represent the minimum and maximum of line items within each suite.
We were not able to run \engine{wasmnow} on all benchmarks; it is clearly fastest on Libsodium and Ostrich.
Besides WasmNow, \engine{sm-base} is the fastest compiler; nearly 3\X faster than the others, and \engine{wazero} is 3\X to 4\X slower than the others.
\OurJit is roughly on par with \engine{v8-liftoff} in compile speed, varying between 0.6\X the speed to 1.5\X the speed over different line items.

To measure code quality of compilers, we compare the execution time of benchmarks relative to \OurJit.
For this experiment, we use a more comprehensive measurement methodology that factors in VM startup and compilation.
If necessary we configure their respective engine to use \emph{only} a specific tier, and disable on-disk caching of compiled code.
Figure~\ref{fig:baseline-execution-time} displays the results of our measurements.

With this data we can approximate each compiler's \emph{SQ-region} (speed-quality region), the general area in the tradeoff space for the runtime of the compiler versus the runtime of the generated code, which is characteristic of the specific compiler.
Figure~\ref{fig:sq-space-baseline} displays the SQ-space for baseline compilers using the same data as Figures~\ref{fig:baseline-execution-time} and \ref{fig:baseline-compile-time}.
It uses a scatter-plot with all benchmark line items to illustrate the variance in both compilation time and execution time across items.
Since many short-running benchmark line items are included, clusters towards the bottom of the graph (lower speedups) indicate where time spent compiling pays off less and VM startup time is more significant.

Our last experiment puts baseline compilers in context with other execution tiers.
We compare baseline compilers to other tiers (interpreters, optimizing JIT compilers, and ahead-of-time translations) in two dimensions: \emph{setup time (S)} and execution speed, or \emph{quickness}.
This makes a larger \emph{SQ-space} that is similar in nature but more general than the compiler \emph{SQ-space} because it includes other setup costs than compiling.
We define \emph{setup time} as the time a VM takes from starting the load of a program to executing its first instruction.
This therefore will characterize the per-module processing time before execution, such as loading and verifying code, building program IR, and compiling.
Since most of these costs are a function of module size, it's reasonable to define their ratio as the \emph{setup speed} and measure it in megabytes per second (\emph{MB/s}).

In this experiment, we measure an even larger set of Wasm execution tiers that includes several interpreters and optimizing compilers, drawn from a larger set of engines.
All new compiler tiers are IR-builders, and all interpreter tiers rewrite the bytecode, with the exception of \OurInt.
Most, but surprisingly not \emph{all}\footnote{\engine{wasm3} does not, in fact, verify the bytecode!}, verify the bytecode.
Thus every engine has some measurable per-module parsing, verification, translation, or compilation cost.
Measuring setup time can be done by instrumenting engines, but requires intrusive modifications.
Instead, we use a simpler, less precise strategy to empirically bound setup time without missing hidden costs.

\newcommand\Mnop{$M_{\texttt{nop}}$}
\newcommand\Mnopm{M_{\texttt{nop}}}

We define $T_{E}(m)$ as the time to execute a module $m$ on engine configuration $E$.
First, we measure VM startup time by executing the smallest possible Wasm module \Mnop, which has only one function that simply returns (total module size is 104 bytes).
We run this hundreds of times to get a statistically significant characterization of startup time.
Next, we approximate the processing cost of each benchmark line-item by inserting an early return in its \wasm{\_start} function, resulting in module $m_0$.
The new module will undergo loading and processing (often compilation) in each engine, but execution time is near zero.

With measurements $T_{E}(\Mnopm)$, $T_{E}(m_0)$, and $T_{E}(m)$:
\begin{itemize}
  \item
    $T_{E}(m_0) - T_{E}(\Mnopm)$ approximates\footnote{In fact, all of these quantities are all subject to sampling error and thus form individual distributions. The resulting ``crude'' approximation is just another distribution that approximates processing time.} the \emph{upper bound} of pre-processing time by removing VM startup time,
  \item $\tilde{T}_{E}(m) = T_{E}(m) - T_{E}(m_0)$ defines the \emph{adjusted execution time} which is the program's execution time without VM startup or module setup time, and
  \item $\tilde{S}_{E,B}(m) = \frac{\tilde{T}_{B}(m)}{\tilde{T}_{E}(m)}$ defines the \emph{adjusted speedup} of configuration $E$ over a baseline config $B$.
\end{itemize}

\subsection{Mapping the Larger SQ-space}
\begin{figure}
  \includegraphics[width=3.5in]{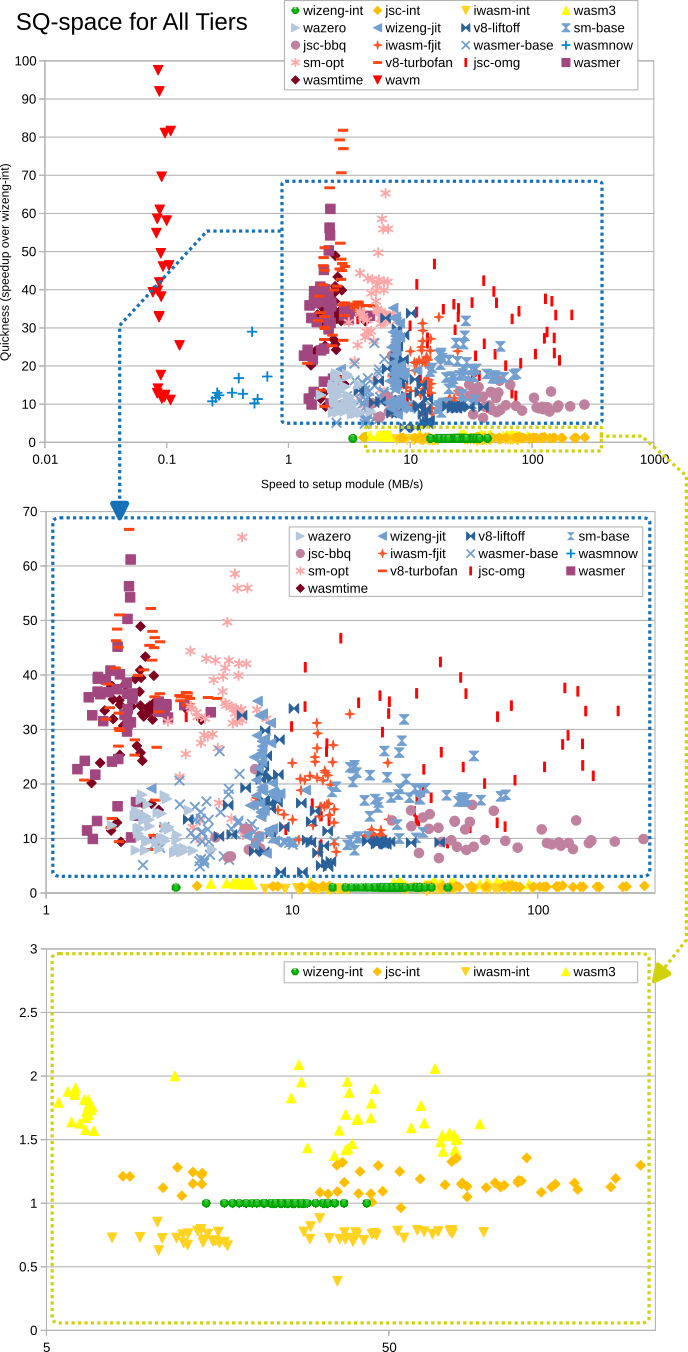}
  \caption{The SQ-space for 18 different Wasm execution strategies.}
  \label{fig:sq-space-all}
\end{figure}

Figure~\ref{fig:sq-space-all} presents averages of 25 runs of each of the 78 benchmark line items on 18 different engines (3 data points each = 106550 data points).
The vertical axis is $\tilde{S}_{E, \text{\engine{wizeng-int}}}(m)$ (i.e. adjusted speedup over \OurInt) and the horizontal axis represents \emph{setup speed}, (the speed of loading, verifying, and translating).
New tiers are:
\begin{itemize}
\item \engine{jsc-int}, \engine{jsc-bbq}, \engine{jsc-omg}, the interpreter, less optimizing, and more optimizing compiler tiers of JavaScriptCore~\cite{JavaScriptCore}.
\item \engine{wasmtime}~\cite{Wasmtime} and \engine{wasmer}, two different Wasm runtimes written in Rust which both use the Cranelift~\cite{Cranelift} optimizing compiler.
\item \engine{wavm}~\cite{WAVM}, a primarily ahead-of-time Wasm engine that uses LLVM.
\item \engine{iwasm-int} and \engine{iwasm-fjit}, the interpreter and fast JIT of the WebAssembly MicroRuntime~\cite{Wamr}.
  \item \engine{wasm3}~\cite{Wasm3}, a fast rewriting interpreter for embedded systems.
\end{itemize}

In the top plot of Figure~\ref{fig:sq-space-all}, we see all tiers compared.
The primarily ahead-of-time \engine{wavm} engine uses LLVM to compile up-front; a slow compiler, this is clearly the slowest at setting up due to a large compile time.
Apparent in the zoomed-in middle plot, baseline compilers (blue colors) all cluster together in the middle; they all have very similar speedups, and though they vary by an order of magnitude in setup speed, are clearly distinguishable from optimizing compilers (red and purple colors) which definitely produce bigger speedups, about 2\X-3\X faster than baseline compilers, though at an order-of-magnitude slower compile speed than baseline.
When we zoom in on interpreters in the bottom plot, it is clear they have a clear performance ceiling; they are all fairly close to each other, within 2\X of \OurInt.
Interpreter setup time varies the most; we attribute this to the fact that 1) some don't verify bytecode, and 2) all the \engine{jsc-*} (JavaScriptCore) tiers use lazy translation, which we could not control.

In general, laziness (i.e. translating a function upon first invocation) is a confounding factor in these measurements, as lazy compile time is not measured in setup time, but attributed to run time, and therefore the adjusted speedup is lower.
As can be seen in the figure, this might be factor for the \engine{jsc-*} compiler tiers, whose speedups appear lower than other optimizing compilers and setup speeds appear faster.
Another confounding factor is parallelism in compilation.
Some engines have fully parallel compilation pipelines and others do not\footnote{Parallel speedup for multiple compiler threads may be greater for optimizing compilers due to longer work units.}.
We chose to leave default threading settings for all engines.
Benchmark modules used in this study are fairly small, so parallel speedup may not be as big of a factor.
A third confounding factor is caching of compiled code.
After noticing anomalies in initial experiments\footnote{Optimizing tiers with instant startup? Too good to be true.} \engine{wasmtime} and \engine{wasmer}, we disabled caching in both of these.

Overall, we can see a great diversity of execution characteristics for Wasm engines, as each tier tends to occupy its own region in this space.
Precision of the plot could probably be improved with metrics reported directly from instrumenting engines.
Nevertheless, we believe the SQ-space analysis provides insight into tradeoffs in a new way and can further inform the design of tomorrow's virtual machines.

\vspace{-12pt}
\section{Related Work}

The first disk format for intermediate code was invented as early as 1968, in the first BCPL compiler's O-Code~\cite{BCPL}.
Prioritizing compiler simplicity and speed above code quality is an old idea that has roots at least as far back as the design of the first Pascal compiler~\cite{WirthStyle} in 1970.
Pascal compilers gave rise to the first widely-used intermediate code format, P-code~\cite{Pcode}, in the mid 1970s, which was still in use as late as 1990~\cite{PcodeModula2}.
P-code was certainly not the last portable low-level code, with others such as TIMI~\cite{AS400}, LLVM bitcode~\cite{LLVMBitcode}, PNaCl~\cite{PNaCl} (itself a variant of LLVM bitcode).
Fast P-code translators might be considered the first baseline compilers.

\B{Dynamic Compilation.}
Over the years, many virtual machines and bytecode formats have been developed, from Smalltalk~\cite{Smalltalk80}, to Java~\cite{JvmSpec}, to the Common Language Runtime (CLR).
The first dynamic compilers were simple, fast, and performed little optimization.
They were often instruction-by-instruction translators, with extremely simple, or even no, register allocation.
They were essentially baseline compilers, but some had IR, e.g. to harness type feedback~\cite{SelfTypeFeedback}.
Later, runtime profiling led to more complex compilers that build and optimize IRs.

\B{Copy \& Patch Code Generation}
Recent work~\cite{CopyAndPatch} on fast compilation using code templates.
The key idea is to use an offline compiler (e.g., LLVM) to generate machine code snippets under various register assignments and with ``holes'' for constants.
When compiling Wasm, an assembler isn't needed; instead, a cache supplies the appropriate snippet for the register assignment at each step of abstract interpretation, patched with appropriate constants.
Our paper evaluated the artifacts of that work, but on a subset of the benchmarks, which did confirm fastest compile speed, but execution time was not better than other baseline compilers.
Correspondence with the authors helped us understand SQ-region in Figure~\ref{fig:sq-space-all}, which is explained by the template generation occuring during VM startup.
One issue with a template-based approach is that the number of templates is combinatoric in the possible abstract values.
\OurJit tracks value tags in its abstract state, which could potentially double or quadruple the number of templates needed.

\B{Synthesizing and Verifying JITs.}
Simple compilers are easier to build, specify, verify, and even synthesize.
Recent work~\cite{SynthJIT} has advanced the generation of \emph{correct} JIT compilers from a specification, which demonstrated a instruction-by-instruction compiler for eBPF running in-kernel with correctness guarantees.
Another approach is to verify the output of the compiler for sandboxing properties, and has been employed for Wasm in~\cite{WasmSandboxing}.

\B{Fast compilers in other domains.}
Many other domains than VMs employ dynamic code generation.
Generating machine code without an intermediate representation has been repeatedly shown to dramatically improve compile speed.
For example, the VCode~\cite{VCode} research system improved on its predecessor, DCG~\cite{DCG} by 35\X.
Simple AST-walking compilers have been deployed in database systems and programmable networks.
Regexes are often implemented with JIT compilers today.
For example, all Web engines use JITs in their regex implementations~\cite{Irregexp}, as well as popular libraries~\cite{PCREJit}.

\nocite{Lightbeam}

\B{Fast compilers cooperate with other tiers.}
Today, many production virtual machines employ multiple compiler tiers.
OpenJDK~\cite{OpenJDK} employs an interpreter and two (tierable) JIT compilers; C2, a highly-optimizing sea-of-nodes compiler, and C1, a faster, SSA-based optimizing compiler.
Web engines continue to evolve, and all employ multiple tiers for both JavaScript and WebAssembly.
The V8 JavaScript engine~\cite{V8} became multi-tier in 2010 when its first optimizing compiler ``Crankshaft''~\cite{CrankShaft} joined its fast AST-walking code generator named ``full codegen''~\cite{FullCodeGen}.
In 2018 V8 replaced both tiers with an interpreter and a new TurboFan~\cite{TurboFan} optimizing compiler, and in 2021 added a baseline compiler ``Sparkplug'' for JavaScript ~\cite{Sparkplug}.
The JavaScriptCore~\cite{JavaScriptCore} virtual machine in Safari employs three different compiler designs, even briefly using LLVM as a top-tier optimizing compiler.

\vspace{-4pt}
  \section{Conclusion}
  This paper captured the core design ideas of baseline compilers for Wasm and documented \NumCompilers implementations, which have appeared nowhere in the literature to date.
  As this paper documents, efficient forward-pass register allocation via abstract interpretation is widespread in single-pass Wasm compilers.
  Examples in this paper illustrate and experiments show that single-pass compilers for Wasm can generate good code very quickly.
  This paper also presented the design of a new, state-of-the-art single-pass compiler, \OurJit with the unique design choice of value tags, which simplifies integration with an in-place interpreter for Wasm and the host garbage collector.
  Measurements show that the overhead of \OurJit's value tag approach is mostly eliminated by optimizations and that the resultant performance is on par with production single-pass compilers.
  Discussion compared and contrasted the \NumCompilers designs and experiments evaluated them on benchmarks, showing that single-pass compilers vary in code quality, primarily due to the differences in modeling constants and register allocation.
  Additional benchmarking data allows us to place all single-pass compilers in a two-dimensional speed-quality tradeoff space (SQ-space) with other available execution tiers for Wasm, including rewriting interpreters and optimizing compilers.
  We find these developments extremely exciting; the explosion of execution strategies for WebAssembly holds great promise to shed new light on long-standing tradeoffs in VM design by studying many diverse engines that all accept a common, well-specified code format.
  
\section*{Acknowledgments}

This work is supported in part by NSF Grant Award \#2148301, as well as funding and support from the Web\-Assembly Research Center.
Thanks to Hannes Payer, Toon Verwaest and Clemens Backes on the V8 team for JIT compiler and tiering discussions.
Thanks to Lars Hansen (formerly Mozilla) for questions on the Spidermonkey baseline compiler design.
Thanks to students Bradley Teo, Yash Anand, and Kazuyuki Takayama, and Elizabeth Gilbert for work on the instrumentation framework in \OurEngine.
Thanks to Heather Miller, Josh Sunshine, Jonathan Aldrich, and Anthony Rowe at CMU.
Thanks to Ulan Degenbaev at DFinity.

\bibliographystyle{IEEETran}
\bibliography{paper}

\end{document}